\documentclass[12pt]{article}
\usepackage{epsfig}
\usepackage{graphicx}
\usepackage{amsmath}

\begin{document}

 \newcommand{\beq}{\begin{equation}}
\newcommand{\eeq}{\end{equation}}
\newcommand{\bea}{\begin{eqnarray}} 
\newcommand{\eea}{\end{eqnarray}}
\newcommand{\beqn}{\begin{eqnarray}}
\newcommand{\eeqn}{\end{eqnarray}}
\newcommand{\beas}{\begin{eqnarray*}}
\newcommand{\eeas}{\end{eqnarray*}}
\newcommand{\defi}{\stackrel{\rm def}{=}}
\newcommand{\non}{\nonumber}
\newcommand{\bquo}{\begin{quote}}
\newcommand{\enqu}{\end{quote}}
\newcommand{\qt}{\tilde q}
\newcommand{\m}{\tilde m}
\newcommand{\trho}{\tilde{\rho}}
\newcommand{\tn}{\tilde{n}}
\newcommand{\tN}{\tilde N}
\newcommand{\gsim}{\lower.7ex\hbox{$\;\stackrel{\textstyle>}{\sim}\;$}}
\newcommand{\lsim}{\lower.7ex\hbox{$\;\stackrel{\textstyle<}{\sim}\;$}}


\def\de{\partial}
\def\Tr{ \hbox{\rm Tr}}
\def\const{\hbox {\rm const.}}  
\def\o{\over}
\def\im{\hbox{\rm Im}}
\def\re{\hbox{\rm Re}}
\def\bra{\langle}\def\ket{\rangle}
\def\Arg{\hbox {\rm Arg}}
\def\Re{\hbox {\rm Re}}
\def\Im{\hbox {\rm Im}}
\def\diag{\hbox{\rm diag}}


\def\QATOPD#1#2#3#4{{#3 \atopwithdelims#1#2 #4}}
\def\stackunder#1#2{\mathrel{\mathop{#2}\limits_{#1}}}
\def\stackreb#1#2{\mathrel{\mathop{#2}\limits_{#1}}}
\def\Tr{{\rm Tr}}
\def\res{{\rm res}}
\def\Bf#1{\mbox{\boldmath $#1$}}
\def\balpha{{\Bf\alpha}}
\def\bbeta{{\Bf\beta}}
\def\bgamma{{\Bf\gamma}}
\def\bnu{{\Bf\nu}}
\def\bmu{{\Bf\mu}}
\def\bphi{{\Bf\phi}}
\def\bPhi{{\Bf\Phi}}
\def\bomega{{\Bf\omega}}
\def\blambda{{\Bf\lambda}}
\def\brho{{\Bf\rho}}
\def\bsigma{{\bfit\sigma}}
\def\bxi{{\Bf\xi}}
\def\bbeta{{\Bf\eta}}
\def\d{\partial}
\def\der#1#2{\frac{\d{#1}}{\d{#2}}}
\def\Im{{\rm Im}}
\def\Re{{\rm Re}}
\def\rank{{\rm rank}}
\def\diag{{\rm diag}}
\def\2{{1\over 2}}
\def\ntwo{${\mathcal N}=2\;$}
\def\nfour{${\mathcal N}=4\;$}
\def\none{${\mathcal N}=1\;$}
\def\ntwot{${\mathcal N}=(2,2)\;$}
\def\ntwoo{${\mathcal N}=(0,2)\;$}
\def\x{\stackrel{\otimes}{,}}

\def\ba{\beq\new\begin{array}{c}}
\def\ea{\end{array}\eeq}
\def\be{\ba}
\def\ee{\ea}
\def\stackreb#1#2{\mathrel{\mathop{#2}\limits_{#1}}}

\def\Tr{{\rm Tr}}
\newcommand{\cpn}{CP$(N-1)\;$}
\newcommand{\wcpn}{wCP$_{N,\tilde{N}}(N_f-1)\;$}
\newcommand{\wcpd}{wCP$_{\tilde{N},N}(N_f-1)\;$}
\newcommand{\vp}{\varphi}
\newcommand{\pt}{\partial}
\newcommand{\ve}{\varepsilon}
\renewcommand{\theequation}{\thesection.\arabic{equation}}

\setcounter{footnote}0

\vfill

\begin{titlepage}

\begin{flushright}
FTPI-MINN-14/33, UMN-TH-3404/14\\
\end{flushright}

\vspace{1mm}

\begin{center}
{  \Large \bf  
{Lessons from supersymmetry: 
\\[2mm]
``Instead-of-Confinement''
Mechanism  
}}

\vspace{5mm}

 {\large \bf    M.~Shifman$^{\,a}$ and \bf A.~Yung$^{\,\,a,b}$}
\end {center}

\begin{center}

$^a${\it  William I. Fine Theoretical Physics Institute,
University of Minnesota,
Minneapolis, MN 55455, USA}\\
$^{b}${\it Petersburg Nuclear Physics Institute, Gatchina, St. Petersburg
188300, Russia
}
\end{center}

\vspace{1mm}

\begin{center}
{\large\bf Abstract}
\end{center}

We review physical scenarios in different vacua of  \ntwo supersymmetric QCD
deformed by the mass term $\mu$ 
for the adjoint matter.  This deformation breaks supersymmetry down to \none and, at large $\mu$, the theory flows to \none QCD.
We focus on  dynamical scenarios which can serve as prototypes
of what we observe in real-world QCD. The so-called $r=N$ vacuum 
is especially promising in this perspective. In this vacuum 
an
``instead-of-confinement'' phase was identified previously,  which is qualitatively close to the
conventional QCD confinement:
the quarks and gauge bosons screened at weak coupling, at strong coupling evolve  into
monopole-antimonopole pairs confined by non-Abelian strings. We review genesis of this picture.

\vspace{2mm}

 To be published in {\sl Quarks 50}, Murray Gell-Mann Festschrift, (World Scientific, Singapore, 2015).

\vspace{2cm}

\end{titlepage}

 \newpage



\section {Introduction }
\label{intro}
\setcounter{equation}{0}

Fifty years ago quarks introduced by Gell-Mann and Zweig \cite{1}, which shortly after became three-colored \cite{2,3}
opened the gate to modern high-energy physics. Discovery of asymptotic freedom in Yang-Mills theories \cite{4} and creation of quantum chromodynamics (QCD) \cite{3,4} completed this process. Gradually it became clear that
a peculiar phenomenon in QCD -- quark (or, more generally, color) confinement presented a novel dynamical phase in field theory with which one had never encountered before. This phase is inherent  to Yang-Mills theories at strong coupling. A possible physical explanation of the confinement phase -- the so called dual Meissner effect -- was suggested in the mid-1970s \cite{mandelstam} but nobody managed to demonstrate it analytically under controllable conditions until much later. 

The advent of supersymmetry changed that. In the mid-1990s  Seiberg and Witten considered ${\mathcal N}=2$
super-Yang-Mills theory with the $SU(2)$ gauge group. In this model a vacuum manifold exists, i.e. a (complex) flat direction. Seiberg and Witten identified \cite{SW1} two points on this manifold in which a monopole or a dyon
become massless. A crucial fact was that supersymmetry allowed them to use analytical extrapolations to give a precise meaning to monopoles and dyons in non-Abelian field theory. Then they showed 
 that a weak deformation of the above model forces the monopoles/dyons to condense, thus triggering
the dual Meissner effect in super-Yang-Mills and ensuing color confinement. The Seiberg-Witten vacua \cite{SW1} are referred to as 
the monopole vacua.

The  Seiberg-Witten solution 
is based on a cascade gauge symmetry breaking. At a high energy scale  the
non-Abelian gauge group is broken down to $U(1)$ (or a  generic  Abelian subgroup in more general models)
by condensation of the adjoint scalar field ${\mathcal A}$. An effective low-energy
theory near the monopole vacuum includes the Abelian gauge fields 
and magnetically charged matter represented by light monopoles or dyons.

The latter condense upon a small ${\mathcal N}=2$ breaking deformation $\mu{\mathcal A}^2$ at a much 
lower scale $\mu$.
Their vacuum expectation values (VEVs) are 
of the order\,\footnote{We introduce a shorthand notation for the dynamical scale parameters
$\Lambda_2$ and $\Lambda_1$, for the scale parameters $ \Lambda_{{\mathcal N}=2}$ and $ \Lambda_{{\mathcal N}=1}$ appearing in ${{\mathcal N}=2}$  and ${{\mathcal N}=1}$ theories,   respectively.}
 of $\mu\Lambda_{2}$.
Simultaneously,
formation of confining color-electric flux tubes (strings) occurs. Their tension is minute being proportional to  
$\mu\Lambda_{2}$.

The Seiberg-Witten solution gave a strong impetus for studies of Yang-Mills dynamics at strong coupling. Another inspiration came from the so-called Seiberg duality \cite{Sdual,IS}. Pairs of completely different ${\mathcal  N}=1$ super-Yang-Mills theories 
(supersymmetric QCD, or SQCD for short) were identified  which were related as follows: one theory from the pair in
the ultraviolet domain (UV), flows to the second theory from the pair in the infrared (IR) limit. 
Of course, this can happen only if the second theory is at strong coupling. 
Formally the Seiberg duality is an extension of the
notion of the  't Hooft anomaly matching \cite{Hooft}. Its roots are deeper, however, and shortly after its discovery  were found to ascend to string theory. 

Before the Seiberg-Witten solution theorists were aware of three basic phases  in gauge theories:
the Higgs, Coulomb and confining regimes. It turned out that the phase structure of super-Yang-Mills theory is much richer. For  instance, contrived matter sectors were shown \cite{Cachazo2} to lead to a number of exotic phases unheard of previously. 
 
The Seiberg duality refers to \none massless 
SQCD. Several years ago we started a project of detailing Seiberg's duality
in various vacua by using the exact Seiberg-Witten solution as an intermediate step in our derivations, see e.g. 
\cite{SYdual,SYN1dual,SYchiral,SYzerovac}. Detailing means that we isolated discrete vacua with specific properties (to be discussed below)
by adjusting quark mass parameters which are present in our \ntwo deformed theory. To pass from \ntwo to \none
we had to consider large rather than small values of the parameter $\mu$ in front of  the deformation term. In these studies we discovered yet a novel phase of super-Yang-Mills which we called ``instead of confinement" \cite{SYrvacua}.
A review of this phase is one  of the main goals of this article. 

Another goal -- discussion of confining strings
which are as close as possible to the conjectured QCD strings -- also requires the large-$\mu$ limit.
Indeed, the Seiberg-Witten  confinement at small $\mu$  is essentially 
Abelian \cite{DS,HSZ,Strassler,VY}, with strings of the Abrikosov type. 
We wanted to construct non-Abelian strings, i.e. those with non-Abelian moduli on their world sheet.
At large $\mu$ adjoint scalars of the Seiberg-Witten model decouple and we then approach 
 \none limit in which the confinement
mechanism is expected to  be 
non-Abelian.  Moreover, it is believed that confinement in   real-world QCD is also non-Abelian.

Thus, there were two reasons why we wanted, starting from the small-$\mu$ limit
to pass to large values of $\mu$ (or, equivalently, the \none limit).
We hoped that in passing  
from \ntwo SQCD via the decoupling of the adjoint scalars, we would be able to see
how the Abelian Seiberg-Witten confinement transforms itself into a regime with non-Abelian confinement.

This  program turned out to be challenging.
The effective (dual) theory which describes low-energy physics 
of $\mu$-deformed \ntwo SQCD below the scale of the adjoint VEVs is in fact
the Abelian gauge theory with light monopoles. It is infrared-free  and stays at weak coupling as long as VEVs of the monopoles (not to be confused with the adjoint VEVs) are small enough. As was mentioned above, these VEVs 
are proportional to
$\mu\Lambda_{2}$; thus,  at small $\mu$ the
condition for week coupling is met. However,  moving toward the desired large-$\mu$
limit breaks  this condition. The effective  theory
goes into a strong coupling regime. The analytic control is 
lost. This was a genuine obstacle.

Recent breakthrough developments allowed us to resolve this problem. 
First, we developed a benchmark model which was slightly different from that of Seiberg and Witten.
In our benchmark model the gauge group is  $U(N)$  (rather than $SU(N)$) 
and the number of quark flavors is $N_f$ where
$(N+1)<N_f<3/2\, N$. In  this version of \ntwo SQCD with the $\mu$ deformation switched on one can identify
a number of the so-called called $r$ vacua which are characterized by a parameter  $r$, 
the number of the condensed (s)quarks in the classical domain of large and generic quark mass parameters
$m_A$ (here $A=1,...,N_f$). Clearly, $r$ cannot exceed $N$, the rank of the gauge group. The monopole vacua (there are $N$ of them) corresponds
to $r=0$. Quarks do not condense in the monopole vacua.

The key observation is as follows \cite{SYzerovac,SYchiral}. In this model there is 
a subset of $r$ vacua (to be referred to as {\em zero vacua})
which can be found at $r<(N_f-N)$.
 In these vacua the gaugino condensate
is parametrically small provided the quark mass parameters are small too. This subset  includes a part of the monopole vacua. 
The smallness of the gaugino condensate ensures that quantum effects are small in the zero vacua,
hence physics can be described in terms of weakly coupled dual theory.

In \cite{SYzerovac,SYchiral} we demonstrated that the Seiberg-Witten confinement,
present in the  zero vacua at small $\mu$,  disappears at large $\mu$. This happens because the scale of 
the gaugino condensate controls the confinement
radius (tensions of the confining strings) in a certain
sector of the dual theory. As we increase $\mu$, confinement 
becomes weaker and weaker in this sector, and eventually quarks 
are ``liberated''. Effective dual theory  has U$(N_f-N)$ gauge group.
This theory is infrared-free and stays at weak coupling at low energies.

It  appears to be in the mixed Coulomb-Higgs phase for quarks. Namely,
$r$ quarks condense, while the U$(N_f-N-r)$ subgroup is  in the Coulomb phase, 
see \cite{SYzerovac,SYchiral} for 
a more rigorous and detailed discussion.
Therefore, the  zero vacua, (in particular, the monopole zero vacua of strongly coupled $\mu$-deformed  SQCD)   are not  good  prototypes for physics in real-world QCD.

Nevertheless, $\mu$-deformed SQCD is a rich theory with a variety
of $r$ vacua with different infrared behaviors. We discovered that certain
other vacua in this theory are more promising in providing us with a prototype for real-world QCD dynamics. This will be described below.

As was mentioned above, the zero vacua  support 
a weakly coupled dual description at large $\mu$ due to the smallness of the gaugino condensate. Another exceptional vacuum is the
$r=N$ vacuum in which the maximal number of quarks condense (in the weakly coupled domain of the large quark masses). In this vacuum the
gaugino condensate is {\em identically zero}. This vacuum also has a
weakly coupled -- the so-called $r$-dual -- description, in the  large-$\mu$
limit \cite{SYN1dual,SYrvacua,SYchiral}.

In this paper  we review infrared dynamics in $\mu$-deformed SQCD focusing mostly on the $r=N$ vacuum.
It is just this vacuum which supports a new phase, namely,  the ``instead-of-confinement'' phase
\cite{SYdual,SYN1dual,SYrvacua,SYchiral}.
In this phase 
 quarks and gauge bosons screened at weak coupling evolve at strong coupling into
monopole-antimonopole pairs confined by non-Abelian strings.

These monopole-antimonopole stringy mesons have ``correct''
(adjoint or singlet) quantum numbers with respect to the global group, in much the same way as
mesons in real-world QCD. Moreover, they lie on the Regge trajectories. Thus, this phase is qualitatively
rather similar to what we observe in  real-world QCD. The role
of QCD constituent quarks is played by monopoles.

The paper is organized as follows. In Sec.~\ref{rNlargexi} we review the
$r=N$ vacuum at weak coupling, i.e. at large $\xi$
(the parameter $\xi$ is defined in Eq. (\ref{xis})). In Sec.~\ref{rdual} we turn to $r$-duality
in this vacuum and the ``instead-of-confinement'' mechanism in the
 small-$\mu$ limit.  In Sec.~\ref{intermediatemu} we discuss the large $\mu$-limit.
 Section~\ref{conclusions} briefly summarizes our results and conclusions.

\section{The {\boldmath{$r=N$}} 
vacuum
at large {\boldmath{$\xi$}} 
}
\label{rNlargexi}
\setcounter{equation}{0}

Our benchmark model reduces to $\!$ \ntwo  SQCD with the $U(N)$ gauge group in the absence
of  $\mu$-deformation.
The matter sector consists of $N_f$ massive quark hypermultiplets. 
 We assume that
$N_f>N+1$ but $N_f<\frac32 N$. 
The latter inequality ensures  that the dual theory is  infrared free. 

This  theory is described in detail in  \cite{SYmon,SYfstr}, see also
 the reviews in \cite{SYrev}.
The field content is as follows. The \ntwo vector multiplet
consists of the  U(1)
gauge field $A_{\mu}$ and the SU$(N)$  gauge field $A^a_{\mu}$,
where $a=1,..., N^2-1$, as well as their Weyl fermion superpartners plus
complex scalar fields $a$, and $a^a$ and their Weyl superpartners, respectively.
These complex scalar fields present the bosonic sector of the {\em adjoint scalars}. 

The matter sector of  the U$(N)$ theory contains
 $N_f$ quark multiplets  which consist
of   the complex scalar fields
$q^{kA}$ and $\tilde{q}_{Ak}$ (squarks) and
their   fermion superpartners --- all in the fundamental representation of 
the SU$(N)$ gauge group.
Here $k=1,..., N$ is the color index
while $A$ is the flavor index, $A=1,..., N_f$.

In addition, as was mentioned, we add the mass term $\mu$ 
for the adjoint scalar superfield  breaking \ntwo supersymmetry down to \none. 
This deformation term
\beq
{\mathcal W}_{{\rm def}}=
  \mu\,{\rm Tr}\,\Phi^2, \qquad \Phi\equiv\frac12\, {\mathcal A} + T^a\, {\mathcal A}^a
\label{msuperpotbr}
\eeq
does not break \ntwo supersymmetry in the small-$\mu$ limit,  see \cite{HSZ,VY,SYmon}, however, at
 large $\mu$ this theory  flows to \none\! SQCD. The fields  ${\mathcal A}$ and ${\mathcal A}^a$ 
in Eq.~(\ref{msuperpotbr}) are  chiral superfields, the ${\mathcal N}=2$
superpartners of the U(1) and SU($N$) gauge bosons.

In this theory we can find  a set of $r$ vacua, where  $r$ is 
the number of condensed (s)quarks in the classical domain of large generic quark masses 
$m_A$ ($A=1,...,N_f$, and $r\le N$). In this review  we will focus on  the $r=N$ vacua. Dynamical scenarios in the $r<N$ vacua are considered in \cite{SYzerovac,SYchiral}. 

These
 vacua have  the maximal possible number of condensed quarks, namely, $r=N$. Moreover,
  the gauge group U$(N)$ is completely 
Higgsed in these vacua, and, as a result, 
they support non-Abelian strings \cite{HT1,ABEKY,SYmon,HT2}.   
These strings result in 
confinement of monopoles. 

First, we  will assume that $\mu$ is small, much smaller than
the quark masses
\beq
|\mu| \ll | m_A |, \qquad A=1, ..., N_f\,.
\label{smallmu}
\eeq
In the quasiclassical domain of large quark masses the squark fields develop VEVs triggered by the 
deformation parameter $\mu$,
\beqn
\langle q^{kA}\rangle &=& \langle\bar{\tilde{q}}^{kA}\rangle=\frac1{\sqrt{2}}\,
\left(
\begin{array}{cccccc}
\sqrt{\xi_1} & \ldots & 0 & 0 & \ldots & 0\\
\ldots & \ldots & \ldots  & \ldots & \ldots & \ldots\\
0 & \ldots & \sqrt{\xi_N} & 0 & \ldots & 0\\
\end{array}
\right),
\nonumber\\[4mm]
k&=&1,..., N\,,\qquad A=1,...,N_f\, ,
\label{qvev}
\eeqn
where we present the squark fields as  matrices in the color ($k$) and flavor ($A$) indices,
 while new parameters $\xi$ are given (in the quasiclassical
approximation) by 
\beq
\xi_P \approx 2\;\mu m_P,
\qquad P=1,...,N\,.
\rule{0mm}{4mm}
\label{xis}
\eeq

The quark condensate (\ref{qvev}) implies  the spontaneous
breaking of both gauge and flavor symmetries.
A diagonal global SU$(N)$ combining the gauge SU$(N)$ and an
SU$(N)$ subgroup of the flavor SU$(N_f)$
survives in the limit of equal (or almost equal)  quark masses. 
This is the color-flavor locking. 

Thus, the unbroken global symmetry we are left with 
is  
\beq
  {\rm SU}(N)_{C+F}\times  {\rm SU}(\tN)\times {\rm U}(1)\,,\qquad \tN\equiv N_f-N\,.
\label{c+f}
\eeq
Here SU$(N)_{C+F}$ is an unbroken global  color-flavor rotation, which involves only the
first $N$ flavors, while the SU$(\tN)$ factor refers to the flavor rotation of the remaining
$\tN$ quarks.

The presence of the global SU$(N)_{C+F}$ symmetry is the reason for
formation of the non-Abelian strings \cite{HT1,ABEKY,SYmon,HT2,SYfstr}.
At small $\mu$ these strings are BPS-saturated \cite{HSZ,VY}, and their
tensions  are determined \cite{SYfstr} by 
the parameters $\xi_P$ introduced in  (\ref{xis}),
\beq
T_P=2\pi|\xi_P|\, , \qquad P=1,...,N.
\label{ten}
\eeq
The above non-Abelian strings, with non-Abelian moduli in the coset
$$\mbox{SU$(N)_{C+F}$/SU$(N-1)\times $U(1)}$$ on their world sheet, 
  confine monopoles. In fact, in the U$(N)$ theories confined  elementary monopoles 
are junctions of two ``neighboring''  strings with labels $P$  and $(P+1)$, see \cite{SYrev} for a more detailed review.

Now, let us briefly discuss the elementary excitation spectrum in the bulk. 
Since
both U(1) and SU($N$) gauge groups are broken by the squark condensation, all
gauge bosons become massive.
To the leading order in $\mu$, \ntwo supersymmetry is not broken.  In fact, with 
nonvanishing $\xi_P$'s (see Eq.~(\ref{xis})), both the quarks and adjoint scalars  
combine  with the gauge bosons to form long \ntwo supermultiplets \cite{VY}. 
In the  equal  quark mass limit  $\xi_P\equiv\xi\,,$  and all states come in 
representations of the unbroken global
 group (\ref{c+f}), namely, in the singlet and adjoint representations
of SU$(N)_{C+F}$,
\beq
(1,\, 1), \quad (N^2-1,\, 1),
\label{onep}
\eeq
 and in the bifundamental representations
\beq
 \quad (\bar{N},\, \tN), \quad
(N,\, \bar{\tN})\,.
\label{twop}
\eeq
The representations in (\ref{onep}) and (\ref{twop})  are marked with respect to two 
non-Abelian factors in (\ref{c+f}). The singlet and adjoint fields are (i) the gauge bosons, and
(ii) the first $N$ flavors of squarks $q^{kP}$ ($P=1,...,N$), together with their 
fermion superpartners.
The bifundamental fields are the quarks $q^{kK}$ with $K=N+1,...,N_f$.
Quarks transform in the two-index representations of the global
group (\ref{c+f}) due to the color-flavor locking. 

The above quasiclassical analysis is valid if the theory is at weak coupling. 
From (\ref{qvev}) we see that the weak coupling condition is 
\beq
\sqrt{\xi} \sim \sqrt{\mu m}\gg\Lambda_{2}\,,
\label{weakcoup}
\eeq
where  we assume all quark masses 
to be of the same order $m_A\sim m$. This condition means that 
the quark masses are large enough to compensate for
the smallness of $\mu$.

\section {\boldmath{$r$}-Dual theory}
\label{rdual}
\setcounter{equation}{0}

In this section we  review non-Abelian $r$ duality in the $r=N$ vacua first established
 in \cite{SYdual,SYtorkink} at small $\mu$. This is an important part of our consideration on which we  
base further analysis, in particular   the conclusion of the instead-of-confinement phase.
 
Let us relax the condition (\ref{weakcoup}) and pass
to the strong coupling domain at 
\beq
|\sqrt{\xi_P}|\ll \Lambda_{2}\,, \qquad | m_{A}|\ll \Lambda_{2}\,,
\label{strcoup}
\eeq
while keeping $\mu$ small.

In non-supersymmetric theories such as QCD this step cannot be carried out 
analytically.
This is the point where supersymmetry becomes important. More exactly,
we exploit the exact Seiberg-Witten solution  on the Coulomb branch \cite{SW1} in our theory. We start at large $\xi\sim \mu m$ (in the equal quark mass limit) and then
go to the equal mass small-$\xi$ limit  via the domain of large 
$\Delta m \sim \Delta m_{AB}\equiv (m_A-m_B)$.

At $\Delta m\sim\Lambda_{2}$ the theory enters a strong coupling regime and undergoes a crossover. We use the Seiberg-Witten curve 
to find the dual gauge group \cite{SYdual,SYN1dual}.
 The  domain (\ref{strcoup}) 
can be described in terms of weakly coupled (infrared free) $r$-dual theory  with  the  gauge group
\beq
{\rm U}(\tN)\times {\rm U}(1)^{N-\tN}\,,
\label{dualgaugegroup}
\eeq
 and $N_f$ flavors of light quark-like dyons.\footnote{The SU$(\tN)\rule{0mm}{4mm}$
 gauge group  was first identified   
at the root of the baryonic Higgs branch in   \ntwo  SU($N$)  SQCD with massless quarks and vanishing  $\xi$ parameters
using the Seiberg-Witten curve in  \cite{APS}.}
Note, that we refer  to our 
dual theory as the ``$r$ dual'' because \ntwo duality described here can be generalized to other
$r$ vacua with $r>N_f/2$.  

This leads to a theory with the dual gauge group U$(N_f-r)\times$U(1)$^{N-N_f+r}$
\cite{SYrvacua}. However, the \none deformation of these $r$ dual theories  
at larger $\mu$ can be performed at weak coupling  only in the $r=N$ 
vacuum \cite{SYzerovac}, which we will discuss below.

\vspace{2mm}

The light dyons $D^{lA}$ 
($l=1, ... ,\tN$ and $A=1, ... , N_f$) are in 
the fundamental representation of the gauge group
SU$(\tN)$ and are charged under the Abelian factors indicated in Eq.~(\ref{dualgaugegroup}).
 In addition, there are  $(N-\tN)$ 
light dyons $D^J$ ($J=\tN+1, ... , N$), neutral under 
the SU$(\tN)$ group, but charged under the
U(1) factors. 

The color charges of all these dyons are identical to those of quarks.\footnote{$\rule{0mm}{4mm}$Because of monodromies \cite{SW1,BF} the quarks pick up at strong coupling root-like color-magnetic 
charges in addition to their weight-like color-electric charges  \cite{SYdual}.}
This is the reason why we call them quark-like dyons. However, 
these dyons are not quarks \cite{SYdual}. As we will show below they belong to a
different representation of the global color-flavor locked group.
Most importantly, condensation of these dyons still 
leads to confinement of {\em monopoles}.

\vspace{2mm} 
The dyon condensates have the form \cite{SYfstr,SYN1dual}:
\beqn
\!\!\!\!
\langle D^{lA}\rangle \!\!\! \!& =& \langle \bar{\tilde{D}}^{lA}\rangle \! =
\frac1{\sqrt{2}}\,\left(
\begin{array}{cccccc}
0 & \ldots & 0 & \sqrt{\xi_{1}} & \ldots & 0\\
\ldots & \ldots & \ldots  & \ldots & \ldots & \ldots\\
0 & \ldots & 0 & 0 & \ldots & \sqrt{\xi_{\tN}}\\
\end{array}
\right)\!,
\label{Dvev}
\\[4mm]
\langle D^{J}\rangle &=& \langle\bar{\tilde{D}}^{J}\rangle=\sqrt{\frac{\xi_J}{2}}, 
\qquad J=(\tN +1), ... , N\,.
\label{adiiz}
\eeqn
The  important feature apparent in (\ref{Dvev}), as compared to the squark VEVs  in the 
original theory (\ref{qvev}),  is a ``vacuum leap'' \cite{SYdual}.
Namely, if we pick up the vacuum with nonvanishing VEVs of the  first $N$ quark flavors
in the original theory at large $\xi$,  and then reduce $\xi$ below 
$\Lambda_{2}$, 
the system undergoes  a crossover transition and ends up in the vacuum 
of the  $r$-dual theory with the dual gauge group (\ref{dualgaugegroup}) and 
nonvanishing VEVs of $\tN $ last dyons (plus VEVs of $(N-\tN)$ dyons 
that are SU$ (\tN)$ singlets).

The parameters $\xi_P$  in (\ref{Dvev}) and (\ref{adiiz}) are determined  \cite{SYfstr} by the 
quantum version of the classical expressions
(\ref{xis}).   
They can be written  in  terms of roots of the Seiberg--Witten curve. 

The first $\tN$ parameters $\xi_P$ which determine VEVs of the non-Abelian dyons in (\ref{Dvev}) are small,
\beq
\xi_P=2\mu\; m_{P+N}\sim \xi^{\rm small} \sim \mu m, \qquad P=1,...,\tN.
\label{xismall}
\eeq
This is a reflection of the fact that the 
non-Abelian sector of the dual theory is infrared free and is at weak coupling
in the domain (\ref{strcoup}). 
Other $\xi$'s which determine VEVs of the Abelian dyons in (\ref{adiiz}) are large,
\beq
\xi_P\sim \xi^{\rm large} \sim \mu \Lambda_{{\mathcal N}=2} , \qquad P=\tN+1,...,N\,.
\label{xilarge}
\eeq

As long as we keep $\xi_P$ and masses small enough (i.e. in the domain (\ref{strcoup}))
the coupling constants of the
infrared-free $r$-dual theory (frozen at the scale of the dyon VEVs) are small:
the $r$-dual theory is at weak coupling.

\subsection{``Instead-of-confinement'' mechanism}
\label{instoc}

Now, we 
are ready to explain the regime which we called ``instead-of-confine\-ment."
Let us  consider the limit of almost equal quark masses.
Both, the gauge group and the global flavor SU($N_f$) group, are
broken in the vacuum. The form of the dyon VEVs in (\ref{Dvev}) shows that the $r$-dual theory 
is also in the color-flavor locked phase. 
 Namely, the  unbroken  global group of the dual
theory is 
\beq
 {\rm SU}(N)\times  {\rm SU}(\tN)_{C+F}\times {\rm U}(1)\,,
\label{c+fd}
\eeq
where this time the SU$(\tN)$ global group arises from   color-flavor locking.

 In much the same way as 
in the original theory, the presence of the global SU$(\tN)_{C+F}$ symmetry
is the  reason behind formation of the non-Abelian strings. Their tensions 
are still given by Eq.~(\ref{ten}),
where the parameters $\xi_P$ are determined by (\ref{xismall}) \cite{SYfstr,SYN1dual}.
These strings still confine monopoles \cite{SYdual}.\footnote{$\rule{0mm}{4mm}$
An explanatory remark regarding
our terminology is in order. Strictly speaking, the  dyons carrying root-like electric charges 
are confined as well. We refer to all such states collectively as to
``monopoles.'' This is to avoid confusion with the quark-like dyons which appear in Eqs.~(\ref{Dvev}) and
(\ref{adiiz}). The
latter dyons carry weight-like electric charges. As was already mentioned, their color charges 
are identical to those of quarks, see \cite{SYdual} for further details.}
  
In the equal-mass limit
the global unbroken symmetry (\ref{c+fd}) of the dual theory at small
$\xi$ coincides with the global group (\ref{c+f}) of the original theory in the
$r=N$ vacuum at large $\xi$.  
However,  this global symmetry is realized in two very distinct ways in the dual pair at hand.
As was already mentioned, the quarks and U($N$) gauge bosons of the original theory at large $\xi$
come in the following representations 
 of the global group (\ref{c+f}):
 $$
 (1,1), \,\, (N^2-1,1), \,\, (\bar{N},\tN), \,\,{\rm and} \,\, (N,\bar{\tN})\,.
 $$
At the same time,  the dyons and U($\tN$) gauge 
bosons of the $r$-dual theory form 
\beq
(1,1),\,\, (1,\tN^2-1),\,\, (N,\bar{\tN}), \,\, {\rm and}\,\,
(\bar{N},\tN)
\label{represd}
\eeq 
representations of (\ref{c+fd}). We see that the
adjoint representations of the color-flavor locked
subgroup are different in two theories. 
     
The quarks and gauge bosons
which form the  adjoint $(N^2-1)$ representation  
of SU($N$) at large $\xi$ and the quark-like dyons and dual gauge bosons which 
form the  adjoint $(\tN^2-1)$ 
representation  of SU($\tN$) at small $\xi$ are, in fact, {\em distinct} states \cite{SYdual}.

Thus, the quark-like dyons are not quarks. 
At large $\xi$ they are heavy solitonic states. However below the crossover
at small $\xi$ they become light and form fundamental ``elementary" states $D^{lA}$ of the $r$-dual theory.
And {\em vice versa}, quarks are light at large $\xi$ but become heavy below the crossover.

This raises the question: what exactly happens to quarks when we reduce $\xi$? 

They are in the ``instead-of-confinement'' phase. The
Higgs-screened quarks and gauge bosons 
at small $\xi$  decay into the monopole-antimonopole 
pairs on the curves of marginal stability (the so-called wall crossing) \cite{SYdual,SYtorkink}. 
The general rule is that the only states that exist at strong coupling inside the curves of marginal stability 
are those which can become massless on the Coulomb branch
\cite{SW1,BF}. For the $r$-dual theory these are light dyons shown in Eq.~(\ref{Dvev}),
gauge bosons of the dual gauge group and monopoles.
 
\begin{figure}
\epsfxsize=6cm
\centerline{\epsfbox{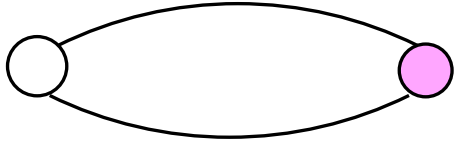}}
\caption{\small Meson formed by a monopole-antimonopole pair connected by two strings.
Open and closed circles denote the monopole and antimonopole, respectively.}
\label{figmeson}
\end{figure}

The
monopoles and antimonopoles produced  at small nonvanishing values of
 $\xi$   in the decay process of the adjoint $(N^2-1,1)$ states
cannot escape from
each other and fly  to opposite infinities 
because they are confined. Therefore, the (screened) quarks and  gauge bosons 
evolve into stringy mesons  (in the strong coupling domain of small  $\xi$) shown in  Fig.~\ref{figmeson},
namely monopole-antimonopole
pairs connected  by two strings \cite{SYdual,SYN1dual}. 

The flavor quantum numbers of stringy  monopole-antimonopole mesons were studied in 
\cite{SYtorkink} in the framework of an appropriate two-dimensional   $CP(N-1)$  model which describes 
world-sheet dynamics of the non-Abelian strings \cite{HT1,ABEKY,SYmon,HT2}. In particular, confined monopoles 
are seen as kinks in this world-sheet theory. If two strings in Fig.~\ref{figmeson} are ``neighboring''
strings $P$ and $P+1$ ($P=1,...,(N-1)$), each meson is in the two-index representation $M_A^B(P,P+1)$
of the flavor group, where the flavor indices are 
$A,B =1,..., N_f$. It splits into a singlet, adjoint and bifundamental representations of 
the global unbroken group (\ref{c+fd}). In particular, at small $\xi$ the adjoint representation
of SU$(N)$  contains former (screened) quarks and gauge bosons of the original theory. 

\begin{figure}
\epsfxsize=6cm
\centerline{\epsfbox{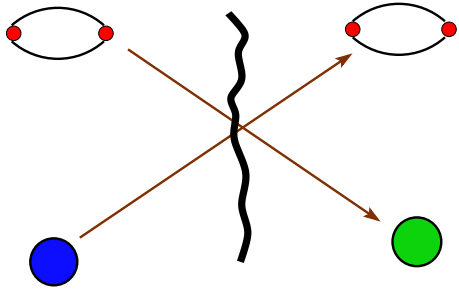}}
\caption{\small Schematic picture of the crossover from large to small 
$\xi$. Blue circles denote light quarks, while green circles denote
light quark-like dyons. Monopole-antimonopole mesons are shown as in Fig.~\ref{figmeson}.}
\label{figlevelcross}
\end{figure}

The picture of the crossover is schematically shown in Fig.~\ref{figlevelcross}. The left and right sides of this figure correspond
to large and small values of $\xi$, respectively. Quarks are light at large $\xi$.
They evolve into monopole-antimonopole stringy mesons at small $\xi$.
Moreover, heavy monopole-antimonopole stringy mesons present at large $\xi$
become light at small $\xi$ and form ``fundamental'' charged matter
of the $r$-dual theory, namely, quark-like dyons.

 We see that the monopole-antimonopole stringy mesons have ``correct''
(adjoint or singlet) quantum numbers with respect to the global group, in much the same way as
mesons in real-world QCD. Let us explain this. For example, in the actual world a $U(1)$ subgroup of 
the global flavor group is gauged with respect to electromagnetic interactions. The correct
(adjoint or singlet) global quantum numbers of 
the monopole-antimonopole stringy mesons mean, in particular, that they have integer 
rather than fractional (like 2/3 or $-1/3$) electric charges. 

Moreover, because these mesons  are formed by strings, they lie on the Regge trajectories. Thus, the monopole-antimonopole mesons
of the instead-of-confinement phase are qualitatively similar to real-world QCD mesons. The role
of QCD constituent quarks is played by monopoles.

\section {Flowing to \boldmath{\none} $\!\!\!\!$  QCD}
\label{intermediatemu}
\setcounter{equation}{0}

In this section we will discuss what happens to the $r$-dual theory in the $r=N$ vacuum 
once we increase $\mu$, see \cite{SYN1dual,SYrvacua,SYchiral}. 
We also discuss the relation of our dual theory to the Seiberg's duality.

\subsection{Intermediate \boldmath{$\mu$}: emergence of the \boldmath{$U(\tN)_{\rm gauge}$} }
\label{UtN}

$\rule{0mm}{5mm}$ Combining Eqs.~(\ref{Dvev}), (\ref{adiiz}), (\ref{xismall}) and (\ref{xilarge}) 
we see that  in the domain
(\ref{strcoup}) the VEVs of the non-Abelian
dyons $D^{lA}$ are  much smaller 
than those of the Abelian dyons $D^{J}$.  
This circumstance is the most crucial. It  allows us to increase $\mu$
and decouple the adjoint fields without violating  the weak coupling condition in the 
dual theory \cite{SYN1dual}.

First let us consider intermediate values of $\mu$   which are large enough to decouple the adjoint matter \cite{SYN1dual,SYrvacua}.
We uplift $\mu$ to the intermediate domain
\beq
|\mu| \gg |m_A |, \qquad A=1, ... , N_f\, , \qquad \mu\ll \Lambda_{2} .
\label{muintermed}
\eeq
The VEVs of the Abelian dyons (\ref{adiiz}) are large. This makes the U(1) gauge 
fields of the dual group 
(\ref{dualgaugegroup}) heavy. Decoupling these gauge factors, together with 
the adjoint matter and 
the Abelian dyons themselves, we obtain the low-energy theory with the $U(\tN)$ gauge fields and a  set of 
non-Abelian dyons 
\beq
D^{lA}\,,\qquad l=1, ... , \tN\,,\quad A=1, ... , N_f\,.
\label{dualgglmurN}
\eeq
The superpotential 
for $D^{lA}$ has the form \cite{SYN1dual}
\beq
{\mathcal W} = -\frac1{2\mu}\,
(\tilde{D}_A D^B)(\tilde{D}_B D^A)  
+m_A\,(\tilde{D}_A D^A)\,,
\label{superpotd}
\eeq
where the color indices are contracted inside each parentheses.
Minimization of this superpotential leads to the VEVs (\ref{Dvev}) for the non-Abelian dyons
determined by  $\xi^{\rm small}$, see (\ref{xismall}).

Below the scale $\mu$, our theory becomes dual to \none SQCD. This $r$-dual theory has  the scale
\beq
\tilde{\Lambda}_{1}^{N-2\tN}= \frac{\Lambda_{2}^{N-\tN}}{\mu^{\tN}}\,.
\label{tildeL}
\eeq
In order to keep this infrared-free theory in the weak coupling 
regime we impose the constraint 
\beq
|\sqrt{\mu m}| \ll \tilde{\Lambda}_{1}\,.
\label{wcdual}
\eeq
This means that at large $\mu$ we must keep the quark masses sufficiently small. 

Note that for the intermediate $\mu$  we assume that 
$\mu\ll \Lambda_{2}$. This condition guarantees that the heavy Abelian $U(1)^{N-\tN}$
sector is at weak coupling too, and is indeed heavy.
If we relax the  condition $\mu\ll \Lambda_{2}$ 
this sector enters a strong coupling regime,
and certain states could in principle become light and show up in our low-energy $U(\tN)$ theory.

\subsection{Connection to Seiberg's duality}
\label{Sconnection}

The gauge group of our $r$-dual theory is U$(\tN)$, the same as the gauge group of the Seiberg's dual
theory \cite{Sdual,IS}.
 This suggests that there should be a close relation between two duals. For intermediate values of $\mu$
this relation was found in \cite{SYvsS,SYzerovac}.

Originally Seiberg's duality was formulated for \none SQCD which in our set-up  corresponds to the limit $\mu\to \infty$.
Therefore, in the original formulation Seiberg's duality referred to the monopole vacua with $r=0$.
Other vacua, with $r\ne 0$, have condensates of $r$ quark flavors  $\langle \tilde{q}q\rangle_{A} \sim \mu m_A$
and, therefore, disappear in the limit  $\mu\to \infty$: they become runaway vacua. 

However,  Seiberg's duality can be (and in fact, was)  generalized to the case of 
$\mu$-deformed \ntwo SQCD  \cite{CKM,GivKut}. If the mass term $\mu$ is large  then $\mu$-deformed \ntwo SQCD flows to \none SQCD with an
additional quartic quark superpotential. This theory has all $r$ vacua which were present in the original \ntwo
theory in the small-$\mu$ limit. 

The generalized Seiberg dual theory in the case of  $\mu$-deformed U$(N)$
\ntwo SQCD at large but finite $\mu$  has the U$(\tN)$ gauge group,
 $N_f$ flavors of Seiberg's ``dual quarks'' $h^{lA}$ (here $l=1,...,\tN$ and $A=1,...,N_f$) and the 
 following superpotential:
\beq
{\cal W}_{S}= -\frac{\kappa^2}{2\mu}\,{\rm Tr}\,(M^2) + \kappa \,m_A\, M_A^A +\tilde{h}_{Al}h^{lB}\,M_B^A,
\label{Ssup}
\eeq
where $M_A^B$ is the Seiberg neutral mesonic field defined as
\beq
(\tilde{q}_A q^B)=\kappa\,M_A^B.
\label{M}
\eeq
The parameter $\kappa$  above has dimension of mass and is needed to formulate Seiberg's duality \cite{Sdual,IS}.
Two last terms in (\ref{Ssup}) were originally suggested by Seiberg, while the first term is a generalization
to finite $\mu$. This generalization originates from the quartic quark potential \cite{CKM,GivKut}.

Now, let us assume the fields $M_A^B$ to be heavy and integrate them out. This implies that $\kappa$ is 
large.
Integrating out the $M$ fields in (\ref{Ssup}) we arrive at
\beq
{\cal W}_{S}^{\rm LE} = \frac{\mu}{2\kappa^2}\, (\tilde{h}_{A}h^{B})(\tilde{h}_{B}h^{A}) + 
\frac{\mu}{\kappa}\,m_A\,(\tilde{h}_{A}h^{A})\,.
\label{SLEsup}
\eeq
The  change of variables
\beq
D^{lA}=\sqrt{-\frac{\mu}{\kappa}}\,\, h^{lA}, \qquad l=1,...,\tN, \qquad A=1,...,N_f
\label{change}
\eeq
brings this superpotential to the form 
\beq
{\cal W}_{S}^{\rm LE} = \frac1{2\mu}\,
(\tilde{D}_A D^B)(\tilde{D}_B D^A)  
-m_A\,(\tilde{D}_A D^A)\,.
\label{S=SY}
\eeq
We see that the $r$-dual and Seiberg's dual theories match each other.  At intermediate $\mu$  the Seiberg $M$ meson is heavy and should be integrated out implying the 
superpotential (\ref{S=SY}) which agrees with the superpotential (\ref{superpotd}) obtained in the $r$-dual 
theory.

This match, together with the identification (\ref{change}), 
reveals the physical nature of Seiberg's ``dual quarks.''
They are not monopoles as one could  naively  think. Instead,
 they are quark-like dyons appearing in the $r$-dual theory below the crossover. Their condensation leads to 
confinement of monopoles and the ``instead-of-confinement'' phase \cite{SYrvacua} for quarks and gauge bosons of the 
original theory.

\subsection {Large \boldmath{$\mu$}}
\label{largemu}

Finally, we pass to the large-$\mu$ domain. Increasing $\mu$ we
simultaneously reduce $m$ keeping $\xi^{\rm small}$  sufficiently small, see (\ref{wcdual}).
Namely, we assume
\beq
\xi^{\rm small}\sim \mu m \ll  \tilde{\Lambda}_{1},
\qquad  \mu \gg \Lambda_{1},
\label{mularge}
\eeq
where  $\Lambda_{1}$ is the scale of the original \none SQCD,
\beq
\Lambda_{1}^{2N-\tN}= \mu^{N}\,\Lambda_{2}^{N-\tN}\,.
\label{Lambda}
\eeq
This ensures that our low-energy U$(\tN)$ $r$-dual theory is at weak coupling. However, the Abelian
U$(1)^{N-\tN}$ sector ultimately enters the strong coupling regime. As was already mentioned, we loose
analytic control over this sector and, in particular, certain states can  become light and show up in 
our low-energy U$(\tN)$ theory. 

This is exactly what happens at large values of $\mu$  and
is, in fact, required by the 't Hooft anomaly matching \cite{Hooft}.
Large values of $\mu$ require a chiral limit of small $m$ due to the condition (\ref{mularge}).
In this limit we need to match global anomalies in terms of original and dual theories.
In fact, without light Seiberg $M$ meson the anomalies do not match.
This was checked initially in \cite{Sdual} for the limit
$\mu \to \infty$ and presented a basis for the discovery of Seiberg's duality. Moreover, recently it was  confirmed \cite{SYchiral} for our $\mu$-deformed
theory at large but finite $\mu$ and massive quarks in the domain (\ref{mularge}).

Natural candidates for the Seiberg  $M$ mesons in the $r$-dual theory are the stringy mesons
$M_A^B(P,P+1)$ (with $P=\tN,...,(N-1)$) from the 
Abelian U(1)$^{N-\tN}$ sector. This sector is at  strong coupling at
large $\mu$; therefore,  certain states from this sector can become light. Perturbative states from this sector
(quark-like dyons and Abelian  gauge fields) are singlets with respect to the global group (\ref{c+fd})
and cannot play the role of the $M$ mesons. Stringy mesons $M_A^B(P,P+1)$ (where $P=1,...,(\tN-1)$)
 from the U$(\tN)$ low-energy theory also cannot play the role of the $M$ mesons. First, they are represented
in the U$(\tN)$ low-energy theory  as nonperturbative solitonic states and cannot be added to 
this theory as new ``fundamental'' or ``elementary" fields. Second, they are too heavy, with masses of the order of
$\sqrt{\xi^{\rm small}}$, determined by the tensions of the non-Abelian strings, which can be calculated  
 at weak coupling.

Thus, we proposed in \cite{SYchiral} that the Seiberg $M_A^B$  mesons  come from a multitude of the
monopole-antimonopole stringy mesons
$M_A^B(P,P+1)$ (where $P=\tN,...,(N-1)$) from the Abelian U(1)$^{N-\tN}$ sector. At large $\mu$ the $M$
meson should become light, with mass of the order of $m$. It should be
incorporated in the U$(\tN)$ low-energy theory   as a new ``fundamental'' or ``elementary" field. 
Note, that other states from the Abelian sector 
are still heavy and decouple.

Since our U$(\tN)$ $r$-dual theory is at weak coupling we can write down its effective action. 
Using
the procedure described in Sec.~\ref{Sconnection} in the opposite direction we ``integrate the $M$-meson
in'' the superpotential (\ref{superpotd}). In this way we arrive at
\beq
{\cal W}= \frac{\kappa^2}{2\mu}\,{\rm Tr}\,(M^2) - \kappa \,m_A\, M_A^A +
\frac{\kappa}{\mu}\,\tilde{D}_{Al}D^{lB}\,M_B^A,
\label{Dsup}
\eeq
where
\beq
\kappa \sim
\left\{
\begin{array}{l}\rule{0mm}{5mm}
 \mu^{\frac34}\Lambda_{{\mathcal N}=2}^{\frac14} \, , \qquad\;  
 \mu\ll \Lambda_{{\mathcal N}=2} \,,\\[4mm]
 \sqrt{\mu m} \, , \qquad \mu\gg \Lambda_{{\mathcal N}=2} \,  .
 \end{array}
 \right.
\label{kappamu}
\eeq
This dependence guarantees that the $M$ meson is heavy, with mass of the order of $\sqrt{\xi^{\rm large}}$ 
at intermediate $\mu$, and becomes light, with mass of the order of $m$ at large $\mu$, see \cite{SYchiral} for details.

\section{Conclusions}
\label{conclusions}

Quarks, gluons, and other notions of which M. Gell-Mann was a pioneer got a new life in the era of supersymmetry,
when supersymmetry-based methods became powerful -- and quite often, unique -- tools in the studies of confinement
and other nontrivial features of gauge dynamics at strong coupling.

\begin{figure}
\epsfxsize=7.5cm
\centerline{\epsfbox{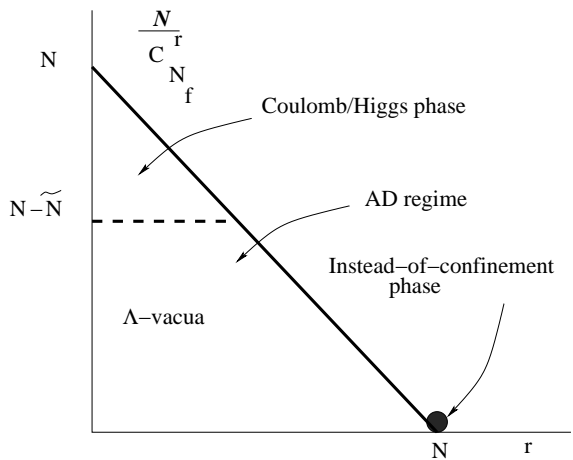}}
\caption{\small Phases of $r$ vacua in \none SQCD. Upper triangle
represents the zero vacua, while black circle denotes the $r=N$ vacuum.}
\label{figphases}
\end{figure}

In this brief article we summarized some applications of non-Abelian strings and
reviewed phases of \none SQCD obtained from $\mu$-deformed \ntwo SQCD in the limit of large $\mu$. 
We identified ``promising" vacua among all $r$ vacua -- promising in the quest of confinement similar to that inherent to QCD.
The number of $r$ vacua as a function of $r$ is shown in Fig.~\ref{figphases}.
 
 The zero vacua represent a subset of vacua at $r<\tN$ with parametrically small gaugino condensate in the limit of the small quark masses. In this limit the zero vacua are described in terms of a weakly coupled dual infrared-free U$(\tN)$
 gauge theory with $r$ condensed quarks. This theory is in the mixed
 Coulomb-Higgs phase \cite{SYzerovac,SYchiral}.
 
The  $\Lambda$ vacua (of which we said little, if at all) have no weak coupling description at large $\mu$ and small $m$ \cite{SYzerovac}.
 In a certain limit they flow into a conformal Argyres-Douglas-like
  strongly coupled regime \cite{AD}.
 
 The regime closest  to what we observe in  real-world QCD
 is represented by the instead-of-confinement phase which occurs in the $r=N$ vacuum at 
 strong coupling. The monopole-antimonopole stringy mesons formed in this
 phase are qualitatively similar to mesons in QCD. They have ``correct''
quantum numbers with respect to the global  group  (singlet plus adjoint) 
 and lie on the Regge trajectories.

\section*{Acknowledgments}
This work  is supported in part by DOE grant DE-SC0011842. 
The work of A.Y. was  supported 
by  FTPI, University of Minnesota, 
by RFBR Grant No. 13-02-00042a 
and by Russian State Grant for 
Scientific Schools RSGSS-657512010.2. The work of A. Y. was supported by 
RSCF Grant No. 14-22-00281.


\vspace{1cm} 

\begin{thebibliography}{99}
\addcontentsline{toc}{section}{References}
\itemsep -2pt

\small

\bibitem{1}
M.~Gell-Mann,
  Phys.\ Lett.\  {\bf 8}, 214 (1964);
G.~Zweig,
{\em An SU(3) model for strong interaction symmetry and its breaking},  CERN-TH-401 (1964), [also in 
  {\sl Developments in the Quark Theory of Hadrons}, Eds. D. Lichtenberg and S. Rosen, (Hadronic Press,
Nonantum, MA, 1980), Volume 1, pp. 22-101].

\bibitem{2}
 O.~W.~Greenberg,
  Phys.\ Rev.\ Lett.\  {\bf 13}, 598 (1964).
  
  \bibitem{3}
 H. Fritzsch and  M. Gell-Mann,	
{\em Current algebra: Quarks and what else? }
in 
Proceedings of the XVI International Conference on High Energy Physics, Chicago,
1972, J. D. Jackson, A. Roberts, eds.) Volume 2, p. 135
[reprinted in hep-ph/0208010]. 

  \bibitem{4}
D.~J.~Gross and F.~Wilczek,
  Phys.\ Rev.\ Lett.\  {\bf 30}, 1343 (1973);
 H.~D.~Politzer,
  Phys.\ Rev.\ Lett.\  {\bf 30}, 1346 (1973).

\bibitem{mandelstam}
Y.~Nambu,
  Phys.\ Rev.\  D {\bf 10}, 4262 (1974);\\
G.~'t Hooft,
{\em Gauge theories with unified weak, electromagnetic and strong interactions,}
in Proc. of the E.P.S. Int. Conf. on High Energy Physics, Palermo, 23-28 June, 1975
ed. A. Zichichi (Editrice Compositori, Bologna, 1976);
Nucl.\ Phys.\ B {\bf 190}, 455 (1981);
S.~Mandelstam,
Phys.\ Rept.\  {\bf 23}, 245 (1976).

    \bibitem{SW1}
N.~Seiberg and E.~Witten,
Nucl. Phys. {\bf B426}, 19 (1994),
(E) {\bf B430},  485 (1994) [hep-th/9407087];
%
Nucl. Phys. {\bf B431}, 484  (1994)
[hep-th/9408099].

\bibitem{Sdual}
  N.~Seiberg,
  Nucl.\ Phys.\  B {\bf 435}, 129 (1995)
  [arXiv:hep-th/9411149].
    
\bibitem{IS}
K.~A.~Intriligator and N.~Seiberg,
  Nucl.\ Phys.\ Proc.\ Suppl.\  {\bf 45BC}, 1 (1996)
  [hep-th/9509066].
  
  
  \bibitem{Hooft}
G.~'t Hooft, {\em Naturalness, Chiral Symmetry, and Spontaneous Chiral Symmetry Breaking},
in 
{\sl Recent Developments In Gauge Theories}
Eds. G.~'t Hooft, C.~Itzykson, A.~Jaffe, H.~Lehmann, P.~K.~Mitter, I.~M.~Singer and R.~Stora,
(Plenum Press, New York,  1980) [Reprinted in {\sl Dynamical Symmetry Breaking}
Ed.  E. Farhi et al. (World Scientific, Singapore, 1982) p. 345
and in G.~'t Hooft, {\sl Under the Spell of the Gauge Principle}, 
(World Scientific, Singapore, 1994), p. 352].
 
 
 \bibitem{Cachazo2}
F.~Cachazo, N.~Seiberg and E.~Witten,
  JHEP {\bf 0304}, 018 (2003)
  [hep-th/0303207].

 \bibitem{SYdual}
  M.~Shifman and A.~Yung,
  Phys.\ Rev.\  D {\bf 79}, 125012 (2009)
[arXiv:0904.1035 [hep-th]].

\bibitem{SYN1dual}
  M.~Shifman and A.~Yung,
  Phys.\ Rev.\  D {\bf 83}, 105021 (2011)
  [arXiv:1103.3471 [hep-th]].
  
\bibitem{SYchiral}
 M.~Shifman and A.~Yung,
  Phys.\ Rev.\ D {\bf 90}, 065014 (2014)
  [arXiv:1403.6086 [hep-th]].
  
  \bibitem{SYzerovac} 
  M.~Shifman and A.~Yung,
  Phys.\ Rev.\ D {\bf 87}, 106009 (2013)
  [arXiv:1304.0822 [hep-th]].
  
  \bibitem{SYrvacua}
 M.~Shifman and A.~Yung,
Phys.\ Rev.\  D {\bf 86}, 025001 (2012)
  arXiv:1204.4165 [hep-th].

\bibitem{DS}
M.~R.~Douglas and S.~H.~Shenker,
Nucl.\ Phys.\ B {\bf 447}, 271 (1995)
[hep-th/9503163].

  \bibitem{HSZ}
A. Hanany, M. Strassler, and A. Zaffaroni,
Nucl.\ Phys.\ B {\bf 513}, 87 (1998)
[hep-th/9707244].

\bibitem{Strassler}
M.~Strassler,
  Prog.\ Theor.\ Phys.\ Suppl.\  {\bf 131}, 439 (1998)
  [hep-lat/9803009].

\bibitem{VY}
A.~I.~Vainshtein and A.~Yung,
Nucl.\ Phys.\ B {\bf 614}, 3 (2001)
[hep-th/0012250].

\bibitem{SYmon}
M.~Shifman and A.~Yung,
Phys.\ Rev.\ D {\bf 70}, 045004 (2004)
[hep-th/0403149].

\bibitem{SYfstr} 
  M.~Shifman and A.~Yung,
  Phys.\ Rev.\ D {\bf 82}, 066006 (2010)
  [arXiv:1005.5264 [hep-th]].
  
   \bibitem{SYrev}
M.~Shifman and A.~Yung,
Rev.\ Mod.\ Phys. {\bf 79}, 1139 (2007),
[arXiv:hep-th/0703267]; an expanded version in {\sl Supersymmetric Solitons,} 
(Cambridge University Press, 2009).

\bibitem{HT1}
A.~Hanany and D.~Tong,
JHEP {\bf 0307}, 037 (2003)
[hep-th/0306150].

\bibitem{ABEKY}
R.~Auzzi, S.~Bolognesi, J.~Evslin, K.~Konishi and A.~Yung,
Nucl.\ Phys.\ B {\bf 673}, 187 (2003)
[hep-th/0307287].


 \bibitem{HT2}
A. Hanany and D. Tong,
JHEP {\bf 0404}, 066 (2004)
[hep-th/0403158].

\bibitem{SYtorkink}
M.~Shifman and A.~Yung,
  Phys.\ Rev.\  D {\bf 81}, 085009 (2010)
  [arXiv:1002.0322 [hep-th]].
  
    \bibitem{APS}
P.~Argyres, M.~Plesser and N.~Seiberg,
Nucl. Phys. {\bf B471}, 159  (1996)
[hep-th/9603042].

  
  \bibitem{BF}
A.~Bilal and F.~Ferrari,
  Nucl.\ Phys.\  B {\bf 516}, 175 (1998)
  [arXiv:hep-th/9706145].
  
    \bibitem{SYvsS}
M.~Shifman and A.~Yung,
Phys.\ Rev.\  D {\bf 86}, 065003 (2012)
[arXiv:1204.4164[hep-th]].

 
  \bibitem{CKM}
G.~Carlino, K.~Konishi and H.~Murayama,
Nucl.\ Phys.\ B {\bf 590}, 37 (2000)
[hep-th/0005076].

   
\bibitem{GivKut}
A.~Giveon and D.~Kutasov,
  Nucl.\ Phys.\ B {\bf 796}, 25 (2008)
  [arXiv:0710.0894 [hep-th]].



\bibitem{AD}
P. C.~Argyres and M. R.~Douglas,
Nucl. \ Phys. \ {\bf B448}, 93 (1995)   
[arXiv:hep-th/9505062];
P. C. Argyres, M. R. Plesser, N. Seiberg, and E. Witten,
Nucl. \ Phys.  \ {\bf B461}, 71 (1996) 
[arXiv:hep-th/9511154].
  
  
  \end{thebibliography}
\end{document}